\documentclass{PoS}

\title{Quasi-spherical accretion in low-luminosity X-ray pulsars: Theory vs.
observations}

\ShortTitle{Quasi-spherical accretion}

\author{\speaker{K.~Postnov}$^a$, N.~Shakura$^a$,  A.~Kochetkova$^a$, L.
Hjalmarsdotter$^a$ \\
    \llap{$^a$}Moscow M.V. Lomonosov State University, Sternberg Astronomical Institute, 13, Universitetskij pr., 119992 Moscow, Russia\\
E-mail: \email{kpostnov@gmail.com}, \email{nikolai.shakura@gmail.com},
\email{apostnova@mail.com}, 
\email{astrogirl@telia.com}}

\abstract{
Quasi-spherical subsonic accretion can be realized in slowly rotating
wind-fed X-ray pulsars at moderate and small X-ray luminosities
$L_x\lesssim 4\times 10^{36}$~erg/s. In this regime the 
accreting matter settles down
subsonically onto the rotating magnetosphere, forming an extended
quasi-static shell. The shell mediates the angular momentum removal 
from the rotating neutron
star magnetosphere by shear turbulent viscosity in the boundary layer
near the magnetosphere or via large-scale convective motions. In the last case
the differential rotation law in the shell is 
close to iso-angular-momentum rotation with $\omega \sim 1/R^2$. 
The
accretion rate through the shell is determined by the ability of the plasma
to enter the magnetosphere due to Rayleigh-Taylor instabilities while taking
cooling into account. Measurements of spin-up/spin-down rates of quasi-spherically wind
accreting X-ray pulsars in equilibrium with known orbital periods (like e.g.
GX 301-2 and Vela X-1) enable determination the main
dimensionless parameters of the model, as well as the estimate of the neutron star 
magnetic field.
For equilibrium pulsars  with independent measurements of the magnetic
field, the velocity of the stellar wind from
the companion can be estimated without the use of complicated spectroscopical measurements.
For non-equilibrium pulsars, there exists a maximum possible spin-down torque 
exerted on the accreting neutron star. From observations of the
spin-down rate and X-ray luminosity in such pulsars (GX 1+4, SXP 1062, 4U
2206+54, etc.) a lower limit on the neutron star magnetic
field is derived, which in all cases turns out to be close to the standard one and in
agreement with cyclotron line measurements.}

\FullConference{An INTEGRAL view of the high-energy sky (the first 10 years)
- 9th INTEGRAL Workshop and celebration of the 10th anniversary of the
- launch,\\
15-19 October 2012\\
Bibliotheque Nationale de France, Paris, France }

\begin{document}



\section{Two regimes of quasi-spherical accretion in X-ray pulsars}

There can be two
different regimes of quasi-spherical accretion (see e.g. 
\cite{Bozzo_ea08} for a recent review of previous studies of wind
 accretion). The captured stellar wind heated up
in the bow shock  at Bondi radius 
$\sim R_B=2GM/v^2$ (where $v$ is the relative stellar wind velocity) 
to high temperatures $k_BT\sim m_p v^2$. If
the characteristic cooling time of plasma $t_{cool}$ is smaller than the
free-fall time $t_{ff}= R_B/\sqrt{2GM/R_B}$, the gas falls
supersonically toward the magnetosphere with the formation of a shock. 
This regime is usually considered in connection with bright XPSRs 
\cite{AronsLea76}, \cite{Burnard_ea83}. The role of X-ray photons
generated near the NS surface is two-fold: first, they heat up plasma in the
bow-shock zone via photoionization, and second, 
they cool down the
hot plasma near the magnetosphere (with $k_BT\sim GM/R_A$) by Compton
processes thus allowing matter to enter the
magnetosphere via the Rayleigh-Taylor instability \cite{ElsnerLamb77}.

In the free-fall accretion regime, the X-ray luminosity (the mass
accretion rate $\dot M$) is determined by the rate of gravitational capture
of stellar wind at $R_B$ (Bondi-Hoyle-Littleton formula, $\dot M\sim \rho v R_B^2$). 
The accretion torque exerted on the NS due to plasma-magnetopshere
interaction is always of  the same sign as the specific angular momentum of
the gravitationally captured stellar wind $j_m$, and the NS can spin-up or spin
down. 

If 
the relative wind velocity $v$ at $R_B$ is slow ($\lesssim 80$~km/s), the  
photoionization heating of plasma is important, but the radiation cooling time of plasma is
shorter than the free-fall time, so the free-fall accretion regime is realized.
If the wind velocity is larger than $\sim 80-100$~km/s, the post-shock
temperature is higher than $5 \times 10^5$~K (the maximum photoionization
temperature for a photon temperature of several keV); for $L_x\lesssim 4\times 10^{36}$~erg/s,
the plasma radiative cooling time is longer than the free-fall time, so 
a hot quasi-spherical shell is formed above the magnetosphere with 
temperature determined by hydrostatic equilibrium \cite{DaviesPringle81},  
\cite{Shakura_ea12}. Accretion of matter through such a shell is subsonic,
so no shock is formed above the magnetosphere. The accretion rate $\dot M$
is now determined by the ability of hot plasma to enter magnetosphere.
This 
is the settling accretion regime. 

\section{Settling accretion regime: theory}

Theory of settling accretion regime was elaborated in \cite{Shakura_ea12}. In this regime, 
the accreting matter subsonically 
settles down onto the rotating magnetosphere forming an extended quasi-static shell.
This shell mediates the angular momentum transfer to/from the rotating NS 
magnetosphere by viscous stresses due to large-scale convective motions and turbulence. 
The settling regime of accretion can be realized for moderate accretion rates 
$\dot M< \dot M_*\simeq 4\times 10^{16}$~g/s. At higher accretion rates, a free-fall gap 
above the neutron star magnetosphere appears due to rapid Compton cooling, and accretion 
becomes highly non-stationary. 

\textbf{Mass accretion rate} through the hot shell is determined by 
mean velocity 
of matter entering the magnetosphere, $u(R_A)=f(u)\sqrt{2GM/R_A}$. The dimensionless factor $f(u)$ is determined by the Compton cooling of plasma above magnetosphere and the critical temperature
for Rayleigh-Taylor instability to develop \cite{ElsnerLamb77}, and is found to be
$f(u)\sim (t_{ff}/t_{cool})^{1/3}$. In the case of Compton cooling
\begin{equation}
f(u)\approx 0.4 \dot M_{16}^{4/11}\mu_{30}^{-1/11}\,,
\end{equation}
where $\dot M_{16}=\dot M/[10^{16} \hbox{g/s}]$ and $\mu_{30}=\mu/[10^{30} \hbox{G}\,\hbox{cm}^3]$
is the NS magnetic moment. The definition of the Alfven radius
in this case is different from the value used for disk accretion: 
\begin{equation}
R_A\approx 1.6\times 10^9[\hbox{cm}]
\left(\frac{\mu_{30}^3}{\dot M_{16}}\right)^{2/11}\,.
\end{equation} 

\textbf{Accretion torques} applied to NS in this regime are determined not only by the specific angular momentum of captured matter $j_m\sim \Omega_b R_B^2$ 
($\Omega_b$ is the orbital angular velocity of the NS), as is the case
of the free-fall accretion, 
but also by the possibility to transfer angular momentum to/from 
the rotating magnetosphere through the shell by large-scale convective motions. 
The plasma-magnetosphere interaction results in emerging of the toroidal magnetic field 
$B_t/B_p=(K_1/\zeta)(\omega_m-\omega^*)/\omega_K(R_A)$, 
where $\omega_m$ is the angular frequency
of matter at the Alfven radus, $K_1\sim 1$ 
the dimensionless coupling coefficient which is different in 
different sources, $\omega_K(R_A)$ is the Keplerian angular frequecy, and 
$\zeta$ is the size of the region of plasma-magnetopshere angular moentum 
coupling in units of the Alfven radius $R_A$. The NS spin evolution equation reads:   
\begin{equation}
\label{sd1}
I\dot\omega^*=\frac{K_1}{\zeta}K_2\frac{\mu^2}{R_A^3}\frac{\omega_m-\omega^*}{\omega_K(R_A)}+
z\dot M R_A^2\,,
\end{equation}
where the second term takes into acount the angular momentum braught to the NS with the
infalling matter ($z<1$). This formula can be rearraged to the form
\begin{equation}
\label{sd_eq}
I\dot \omega^*=Z \dot M R_A^2(\omega_m-\omega^*)+z \dot M R_A^2\omega^*\,,
\end{equation}
where the coupling coefficient is $Z\approx 0.36 (K_1/\zeta)\dot M_{16}^{-4/11}\mu_{30}^{1/11}$.

The gas-dynamical treatment of the problem of angular momentum transfer through the shell by viscous turbulence stresses \cite{Shakura_ea12} showed that $\omega_m\approx \Omega_b(R_A/R_B)^n$, where the index 
$n$ depends on the character of turbulence in the shell. For example, in the case of isotropic near-sonic turbulence we obtain $n\simeq 3/2$, i.e. quasi-Keplerian rotation distribution. In the more likely case of strongly anisotropic turbulence (because of strong convection) we find $n\approx 2$, i.e. an iso-angular-momentum distribution\footnote{If there is no convection in the shell, 
the magnetosphere interacts with the shell in a turbulent boundary layer. In that case
the spin-down torque is $\sim \mu^2/R_c^3$ ($R_c=(GM/\omega*^2)^{1/3}$ is the corotation radius)
and is independent on $\dot M$. This case of weak coupling can be realized for very faint slowly rotating XPSRs.}. 
\section{Settling accretion regime: observations}

\textbf{Equilibrium X-ray pulsars}. In equilibrium XPSRs $<\dot\omega^*>=0$ (e.g. 
Vela X-1 and GX 301-2). In this case, measurements of spin-up/spin-down near the 
equilibrium pulsar period $P_{eq}^*$ (or frequency $\omega^*_{eq}$) provides additional 
quantity $\partial \omega^*/\partial\dot M$ (or $\partial \omega^*/\partial y$, 
where $y\equiv \dot M\dot M_{eq}$ is mass accretion rate or X-ray luminosity noprmalized
to the equilibrium value). In this case (see \cite{Shakura_ea12}, \cite{Shakura_ea13} for more details) for the convective shell ($n=2$) we find:
a) equilibrium NS period via binary orbital period $P_b$, mass
accretion rate $\dot M$ (or X-ray luminosity $L_x=0.1\dot M c^2$), NS magnetic field $\mu_{30}$
and relative stellar wind velocity $v_8\equiv v/(1000 \hbox{km/s})$
\begin{equation}
\label{Peq}
P_{eq}^*\approx 1300 [\hbox{s}]\mu_{30}^{12/11}(P_b/10\hbox{d})
\dot M_{16}^{-4/11}v_8^4\,;
\end{equation}
b) estimate of the coupling parameters $Z_{eq}$ or $(K_1/\zeta)$ via $P^*$ and $\partial \omega^*/\partial y$:
\begin{equation}
\label{Zeqrho}
Z_{eq}\approx\frac{I\frac{\partial \dot\omega^*}{\partial \dot M}|_{eq}}{\frac{4}{11}\omega^*R_A^2}\approx 1.8\left(\frac{\partial \dot\omega^*/\partial y|_{y=1}}{10^{-12}\hbox{rad/s}}\right)(P^*/100s)\dot M_{16}^{-7/11}\mu_{30}^{-12/11}\,;
\end{equation}
c) estimate of the NS magnetic field via $P^*$ and $\partial \omega^*/\partial y$
\begin{equation}
\label{mueqnew}
\mu_{30,eq}\approx 5 \left(\frac{\partial \dot\omega^*/\partial y|_{y=1}}{10^{-12}\hbox{rad/s}}\right)(P^*/100s)\left(\frac{K_1}{\zeta}\right)^{-1}\dot M_{16}^{-3/11}\,.
\end{equation}
d) estimate of the stellar wind velocity 
\begin{equation}
\label{e:v8min}
v_8\approx 0.53 (1-z/Z_{eq})^{-1/4}
\dot M_{16}^{1/11}\mu_{30,eq}^{-3/11}
\left(\frac{P_*/100 \hbox{s}}{P_b/10 \hbox{d}}\right)^{1/4}
\end{equation}
(note here the weak dependence on $\dot M$ and $\mu$). The observed and calculated parameters of the equilibrium wind-fed pulsars Vela X-1 and GX 301-2 are listed in Table 1. Note close values of the coupling parameter $Z_{eq}\sim 3$ (or $\zeta\sim 1/10$) for both pulsars, and the independent 
measurement of the stellar wind velocity similar to the observed values. 

\textbf{Non-equilibrium X-ray pulsars}. From Eq. (\ref{sd1}) it can be shown that $\dot \omega^*$ as a function of $\dot M$ reaches a mimimum at some accretion rate 
$y_{cr}=\dot M_{cr}/\dot M_{eq}=(3/2n+3)^{11/2n}<1$. For $n=2$ we find:
\begin{equation}
\label{e:omegadotsdmax}
\dot\omega^*_{sd,max}\approx -1.13\times 10^{-12}[\hbox{rad/s}] (1-z/Z)^{7/4} \left(\frac{K_1}{\zeta}\right) 
\mu_{30}^{2}v_8^3\left(\frac{P^*}{100\hbox{s}}\right)^{-7/4}
\left(\frac{P_b}{10\hbox{d}}\right)^{3/4}\,.
\end{equation}
At $y<y_{cr}$ the spin-down torque should anti-correlate with the the X-ray flux variations,
$\partial \dot\omega^*/\partial y<0$, with $\dot\omega_{sd}\sim -R_A^{-3}\sim -\dot M^{6/11}$.
This is the case observed in long-term spinning-down 
XPSR GX 1+4 \cite{Chakrabarty_ea97}, \cite{GonzalezGalan_ea12}. From the condition 
 $|\dot \omega^*_{sd}|\le |\dot\omega^*_{sd,max}|$ we obtain the lower limit of the NS magnetic field:
\begin{equation}
\label{e:mulim1}
\mu_{30}>\mu_{30, min}'\approx 0.94
\left|\frac{\dot\omega^*_{sd}}{10^{-12}\hbox{rad/s}}\right| 
\left(\frac{K_1}{\zeta}\right)^{-1/2}
v_8^{-3/2}
\left(\frac{P^*}{100\hbox{s}}\right)^{7/8}
\left(\frac{P_b}{10\hbox{d}}\right)^{-3/8}.
\end{equation}
If spin-up torque can be neglected, we find another estimate of the 
lower limit to the NS magnetic field 
\begin{equation}
\label{e:mulim2}
\mu_{30}>\mu_{30, min}''\approx 1.66 
\left|\frac{\dot\omega^*_{sd}}{10^{-12}\hbox{rad/s}}\right|^{11/13} 
\left(\frac{K_1}{\zeta}\right)^{-11/13}
\dot M_{16}^{-3/13}
\left(\frac{P^*}{100\hbox{s}}\right)^{11/13}.
\end{equation}
Note that in contrast to Eq. (\ref{e:mulim1}), 
this estimate is independent of the poorly known stellar wind velocity $v_8$ and
binary orbital period $P_b$.

With decreasing $\dot M$ in non-equilibrium pulsars, the ratio of the toroidal to poloidal magnetic field components increases, reaching $B_t\sim B_p$ at $\dot M^*_{16}\approx 0.27  \left|\frac{\dot\omega^*_{sd}}{10^{-12}\hbox{rad/s}}\right|^{11/6}\mu_{30}^{-2/3}$. Below this luminosity  
acrretion becomes more non-stationary (likely the case of GX 1+4), but it is not still centrifugally
prohibited. The propeller stage begins once $R_A>R_c$ at much smaller luminosities:
$\dot M^{**}_{16}\approx 0.008 \mu_{30}^{3}
(P^*/100\hbox{s})^{-11/3}$.
\begin{table*}
\label{T2}
 \centering
 \caption{Parameters of pulsars discussed}
 $$
\begin{array}{lcc|ccc}
\hline
\hbox{Pulsars}&\multicolumn{2}{c}{\hbox{Equilibrium}}&
\multicolumn{3}{c}{\hbox{Non-equilibrium}}\\
\hline
& {\rm GX 301-2} & {\rm Vela X-1}
& {\rm GX 1+4} &{\rm SXP1062}&{\rm 4U 2206+54}\\
\hbox{Ref.}& \cite{Doroshenko_ea10} & \cite{Doroshenko11}& \cite{GonzalezGalan_ea12}
& \cite{Haberl_ea12} &\cite{Reig_ea12}\\
\hline
\multicolumn{5}{c}{\hbox{Measured parameters}}\\
\hline
P^*{\hbox{(s)}} & 680 & 283 & 140 & 1062 &5560\\
P_B {\hbox{(d)}} & 41.5 & 8.96 & 1161 & \sim 300^\dag& 19\\
v_{w} {\hbox{(km/s)}} & 300 & 700 & 200 & \sim 300^\ddag& 350\\
\mu_{30}& 2.7 & 1.2 & ? & ? & 1.7\\
\dot M_{16} & 3 & 3 & 1 & 0.6 & 0.2\\
\frac{\partial \dot \omega}{\partial y} \arrowvert_{y=1}{\hbox{(rad/s)}}
& 1.5\cdot10^{-12} & 1.2\cdot10^{-12} & n/a & n/a & n/a\\
 \dot\omega^*_{sd} & 0 & 0 & - 2.34 \cdot 10^{-11} & - 1.63 \cdot 10^{-11} &
-9.4 \cdot 10^{-14}\\
\hline
\multicolumn{5}{c}{\hbox{Obtained parameters}}\\
\hline
f(u) & 0.53 & 0.57 \\
\frac{K_1}{\zeta}& 14 &10& & & \gtrsim 8\\
Z& 3.7 & 2.6\\
B_t/B_p & 0.17 & 0.22\\
R_A{\hbox{(cm)}}& 2\cdot 10^9 & 1.4\cdot 10^9\\
\omega^*/\omega_K(R_A)& 0.07 & 0.08\\
v_{w,min} (km/s)& 500 & 740\\
\mu_{30,min}& & &\mu_{min}'\approx4&\mu_{min}''\approx20&\mu_{min}'\approx
3.6 \\
\hline
\end{array}
$$
\noindent$^\dag$ Estimated from the Corbet diagram\\
\noindent$^\ddag$ Typical velocity assumed in Be X-ray binaries
\end{table*}

\textbf{Very slowly rotating XPSRs} There are known several very slowly rotating XPSRs, including some 
in HMXB (SXP 1062 with $P^*=1062$~s \cite{Haberl_ea12}, 4U 2206+54 with $P^*=5550$~s \cite{Reig_ea12}) and some in 
LMXB (e.g. 3A 1954+319, $P^*=19400$~s \cite{Marcu_ea11}). 
Assuming disk accretion in such pulsars would 
require incredibly high NS magnetic fields (see e.g. discussion in \cite{Wang_12}). 
However, application of our model to these and other non-equilibrium XPSRs (see Table 1) gives the low limits of the NS surface magnetic field 
in the usual range $10^{12}-10^{13}$~G, and it is too preliminary 
to classify these objects as accreting magnetars. 
Note also that if at small X-ray luminosities 
convection is not developed in the shell, a quasi-Keplerian rotation law with $n=3/2$ can be established. The equilibrium NS spin period in this case is $P_{eq}^{(n=3/2)} =
P_{eq}^{(n=2)}\sqrt{R_B/R_A}\sim 10 P_{eq}^{(n=2)}$. That means that NS periods in such XPSRs can 
easily reach a few 10000 s for the standard NS magnetic field.   

\textbf{Other applications}
A possible implication of the settling accretion theory can be for non-stationary 
phenomena in XPSRs. The theory explains the observed temporary 
'off'-states in Vela X-1, GX 301-2,
4U 1907+09, when the plasma cooling near the magnetospheric equator 
occurs due to radiative processes \cite{Shakura_ea12b}. The Compton cooling turns out to be ineffective due to 
X-ray pattern changing from fan-beam to the pencil-beam (as suggested by the observed
X-ray pulse shape changes during the off-state in Vela X-1 \cite{Doroshenko_ea11}). 

A dynamical instability of the shell on 
the time scale of the order of the free-fall time from the magnetosphere 
can appear due to increased Compton cooling
and hence increased mass accretion rate in the shell, leading to  
an X-ray outburst with duration lasting about 
the free-fall time scale of the entire shell
($\sim 1000$~s). Such a transient behaviour 
is observed in supergiant fast X-ray transients (SFXTs) (see e.g. \cite{Ducci_ea10}), 
in which slow X-ray pulsations are found (e.g. IGRJ16418-4532, $P^*\approx 1212$~s
\cite{Sidoli_ea12}). The observed flaring 
behavior can be the manifestation of a 
Rayleigh-Taylor instability from the magnetospheric radius occurring 
on the dynamical time scale $\sim R_A^{3/2}/\sqrt{GM}$.

\section{Conclusion}
        
At X-ray luminosities $<4\times 10^{36}$~erg/s wind-fed
X-ray pulsars can be at the stage of subsonic
settling accretion. In this regime, accretion rate
onto NS is determined by the ability of plasma
to enter magnetosphere via Rayliegh-Taylor instability. 
The angular momentum can be
transferred through the quasi-static shell via
large-scale convective motions initiating turbulence cascade. 
The theory explains long-term spin-down in wind-
fed accreting pulsars and properties of short-term
torque-luminosity correlations. Long-period low-luminosity X-ray pulsars are most
likely experiencing settling accretion too. 
Spectral and
timing measurements of slowly rotating 
X-ray pulsars can be used to further test this accretion regime.

\end{document}